\documentstyle[aps,psfig]{revtex}

\begin{document}
\title{Experimental evidence for narrow baryons in the mass range
1.0$\le$~M~$\le$1.46~GeV}
\author{
B. Tatischeff$^{1}$
\thanks{Corresponding author: B. Tatischeff, \vspace*{7.mm}E-mail:
tati@ipno.in2p3.fr}, J. Yonnet$^{1}$,
M. Boivin$^{2}$, M. P. Comets$^{1}$, P. Courtat$^{1}$,\\
R. Gacougnolle$^{1}$, Y. Le Bornec$^{1}$,
E. Loireleux$^{1}$, F. Reide$^{1}$,\\ and N. Willis$^{1}$}
\date{ }
\address{
$^1$Institut de Physique Nucl\'eaire,CNRS/IN2P3, F--91406 Orsay Cedex,
France\\
$^2$Laboratoire National Saturne,CNRS/IN2P3, F--91191
Gif-sur-Yvette Cedex, France}

\maketitle
\vspace*{1cm}
\begin{abstract}
The reaction p~p $\rightarrow$ p ${\pi^+}X$ was studied at different incident
energies around T$_{p}$=2~GeV. Narrow baryonic structures were
observed in the
missing mass M$_{X}$ and in the invariant mass M$_{p\pi^{+}}$.
 The masses of
these structures are: 1004, 1044, 1094, 1136, 1173, 1249, 1277, and 1384~MeV
(and possibly 1339~MeV). Some of them 
were also observed at the same masses in the missing
mass spectra of the d~p~$\rightarrow$~p~p~X reaction although with a weaker
signature. Many checks were performed to make sure
that these structures were not produced by experimental artifacts.
Several narrow
small amplitude peaks, were also extracted using already published
photo-nucleon
cross sections. 
The small widths of all these results, and the stability of the 
observed structures, regardless of the experiment, were used to conclude that
they are genuine baryons which and not merely the consequence of
dynamical rescatterings.
These baryons cannot be associated with classical $q^{3}$ quark configurations.
We associate them with two colored cluster quark configurations.\\
\end{abstract}
  
  
\section{INTRODUCTION}
For a number of years narrow structures have been observed in hadronic spectra.
Isovectorial dibaryons were first observed \cite{bor}
(and references therein), then other narrow dibaryon isospin
states \cite{tro}.
Narrow structures  were also observed later in baryons \cite{bor1} and in
mesons \cite{yon} \cite{bor4} \cite{tro1}. Since these structures were
observed at stable masses, whatever the reaction, incident energy or
production angle, they were associated with real hadrons and not with final
state rescatterings. There is no room
for new low mass mesons within the many quark models for mesons, that
consider only
$q\bar{q}$ configurations,
given the excellent agreement between observation and calculations \cite{kol}.
Once again there is no room in
the mass range M$\le$1.5~GeV for new baryons within the many quark
models for baryons, if we
consider only $qqq$ configurations. The classical baryonic spectrum, which
corresponds to broad resonances, is rather
well understood. This is not the case for the narrow structures observed.\\
\hspace*{4.mm}An interpretation was tentatively proposed to explain these
narrow and exotic hadrons, associating them with two colored quark clusters.
It was shown in the already quoted references that such an attempt allows to
reproduce the masses with models using a very small number of 
adjusted parameters (from zero
to two), depending on the hadronic species in question.
In these models, where the exotic hadron arises from a partial deconfinement,
we would expect small narrow amplitude signatures throughout the complete
hadronic spectra. This paper presents the
analysis and the results obtained in order to check this idea in low mass
baryons. Preliminary results concerning the three first exotic baryons 
have previously been published \cite{bor1}.\\
\hspace*{4.mm}This paper is constructed in the following way: in Sec II
we will describe the detector performances, and 
the data processing; in Sec. III the various normalizations
applied in order to get the cross sections will be described; the simulation
used to make 
some of these normalizations and to test the complete experimental device
will be discussed in Sec. IV whereas the various checks performed in order to
underline our confidence in the genuine reality of the observed structures
are presented in Section V. In
Section VI  the results of our two experiments are presented, namely:\\
\hspace*{2.mm}- the p~p $\rightarrow$ p ${\pi^+}X$ reaction, studied at
three incident energies: T$_{p}$=1.52 GeV, 1.805 GeV, and 2.1~GeV and several
forward angles. Narrow baryons were looked for in the missing mass
M$_{X}$ and in the invariant mass M$_{p\pi^{+}}$;\\
\hspace*{2.mm}- the missing mass of the
$\vec{\mbox{d}}$~p~$\to$~p~p~X reaction was
studied at two incident deuteron energies: T$_{d}$=1.722 GeV and 2.1~GeV.\\
\hspace*{4.mm}We compare the masses of our structures to the masses
of small structures which can be extracted from various reactions studied 
with incident hadrons or leptons by different collaborations. The
observed masses will be discussed in section VII within a phenomenological
model based on two quark cluster configurations. Finally, the last part is
devoted to conclusions.
\section{EXPERIMENT}
\subsection{Experimental description}
The measurements were performed  at the Saturne synchrotron beam facility. The
proton beam energies were 1.52 GeV, 1.805 GeV and 2.1~GeV, and
the deuteron beam energies were 1.72 GeV and 2.1~GeV.
The particles were detected using the SPES3 spectrometer and detection system
(see Fig. 1). The main properties of the spectrometer were:\\
 - a solid angle of $\pm$~50~mrad in both the
horizontal and vertical planes, and\\
 - a large momentum
range (600 MeV$<$pc$<$1400~MeV) at B=3.07~T.\\ 
\hspace*{4.mm}The liquid H$_{2}$ target of 393~mg/cm$^{2}$ was held in a
container with 130~$\mu$m thick Ti windows. External heat shields comprised
 of 24~$\mu$m thick aluminium were placed in the beam-line on either side of
the target. The beam particles/burst varied between $10^{8}$/burst and
5x$10^{8}$/burst, depending on the spectrometer angle
and the incident energy. It was chosen in such a way that the acquisition
dead time was less than
10\%.  The vertical
angular acceptance was
determined using the spectrometer magnetic field map and GEANT code 
\cite{gea}. The drift
chambers, the trigger and the acquisition code were
conceived in order to detect, to identify and to measure
the properties of two-particle reactions.\\
\hspace*{4.mm}Several drift chambers were used to reconstruct the particle
trajectories. The first chamber C1 \cite{ber} (called ``MIT-type''),
was situated on
the spectrometer focal plane. Its spatial and angular horizontal
 resolutions are~: $\sigma_x$=90~$\mu$m and
 $\sigma_{\theta}$=18~mrad respectively. Its efficiency was carefully
 calibrated using the other drift chambers. 
Two multidrift chambers, C2-C3 (called the ``CERN-type''
 chambers), placed perpendicular to the average particle direction, were
designed to get information on trajectories in the horizontal and
vertical planes. However, due to the small vertical magnification of the
spectrometer ($\approx$~0.14),
the $\phi$ resolution at the target was too poor to be useful.\\
\hspace*{4.mm} The trigger consisted of four planes of plastic scintillator
 hodoscopes. The first and last planes were each made
 of 20 scintillators each. The dimensions of each plastic detector were
 12$\times$40~cm$^2$ for the first plane (A), and
 18$\times$80~cm$^2$ for the last plane (B).  The
 time of flight baseline from the first scintillator plane to the last
 one was 3~m. Particles were identified by their time of
  flight
  between the A$_i$ and B$_j$ detectors and also by their energy loss
  in the A$_i$ detectors. This latter measurement was mainly used to
  discriminate between one and two charged particles. Mean-timers and
  constant fraction discriminators were used and the time resolution for
  each scintillator was typically $\sigma$~=~180~ps.
  The large horizontal
  angular magnification of the spectrometer
  produced a large horizontal angular opening (up to $30^0$)
  of the trajectories at the
  output of the spectrometer. This resulted in
  a large number of useful A$_i$B$_j$ combinations (125), between
  the first and last scintillator planes, which required
  the same number (125) coincidences. It is important to note that a
  mean range of $\approx$~200~MeV/c ($25\%$ of the focal plane
 momentum dependant acceptance) is seen by each A$_i$B$_j$ combination.
 Therefore, there is a large overlap between many A$_i$B$_j$ trigger
 combinations for each spectrometer momentum.
 The count rate at each momentum value is then less liable to be influenced
 by errors from the scintillator efficiency.  Careful trigger
 efficiency measurements of all the 125
   combinations  were performed using a system of scintillator counters
   moving in front of the A-hodoscope and behind the B-hodoscope.
   The mean value of the trigger efficiency was $\approx~95\%$.\\
\hspace*{4.mm}When the scattering angle or the incident energy
vary, the mean value of the missing mass M$_{X}$ moves for a
given A$_i$B$_j$ coincidence.\\
\subsection{MIT drift chamber efficiency}
The MIT drift chamber was located in the focal plane of the spectrometer.
It therefore played a major role in our detection system, justifying a careful
efficiency calibration. The efficiency of the MIT drift chamber
was evaluated using the CERN drift
chambers, by calculating the ratio of the number of times where all three
detectors were hit to the number where only the two CERN chambers were hit.
Figure 2 shows that this efficiency displays
a slowly varying value as a function of position. 
Consequently this efficiency correction cannot be a source of eventual
structures.
In Fig. 2 the circles correspond to
events where two particles were detected, and the triangles to single
particle events. Thanks to the large
momentum bite of the spectrometer, the data were taken
without having to move the
magnetic field. Consequently, the regularly spaced structures seen in the 
results cannot be
attributed to an eventual mis-treatment of the MIT drift chamber efficiency.\\
\subsection{Pion-proton identification}
In Fig.~3, four typical time of flight raw spectra,
chosen among the 125 different possibilities, are shown.
 We observe the quality of the particle
identification by looking at the two peak separation corresponding to pion
and proton times of flight. The four inserts correspond to different
spectrometer mean momenta and different angles.
The inserts (a), (b), (c), and (d) correspond respectively to 1.36~GeV/c (top
 spectrometer momentum value), 1.25~GeV/c, 1.00~GeV/c, and 0.83~GeV/c.
 The separation is
always very good, even on the high momentum side.\\
\hspace*{4.mm}A second level trigger filters the events that are stored.
This was done thanks to a second
time of flight between both detected particles which was used
online in order to control the intensity dependent random hit rate. The same
device was also used offline in order
to reject the very small
amount of misidentified particles between pions and protons at the top of
the momentum range (see Fig. 3(a)). The time measured was corrected using
the particle momenta, the corrections being calculated from the variable
distances from the target to the first
scintillating plastic hodoscope plane~A.
After pedestal corrections and variable distance corrections, the time
differences of all
the $A_{i}A_{j}$ time of flights (20$\times$(20-1)/2=190 combinations) 
were plotted on the same histogram and are shown in
 Fig.~4. We observe a tail due to accidental coincidences to the left of the
 peak. A window of $\pm$~9~channels, as drawn on the figure, was used in the
 detailed data analysis. Incorrect proton-pion identifications fell
 around channel
 -21, and can be seen as a small peak in the figure. These  misidentified
 proton-pion events correspond to events that lie in the tail crossovers of
 the particle identification at the top end of the momentum scale 
 (see Fig. 3(a)). The spectra of the first time of flight (Fig. 3) were vetoed
 by the second time of flight.\\ 
\subsection{Mass resolution}
An important property of the experiment is the good mass resolution,
essential when narrow and weakly excited structures are being looked for.
The resolution in the invariant mass was measured using the final
state interaction peak (FSI) from the p~p~$\to$~p~p~X reaction.
Figure 5 shows these distributions for all three energies, vertically
shifted in order to keep the figure easy to read. Here, full circles,
squares, and triangles correspond respectively to results from
T$_{p}$=1.52~GeV, 1.805~GeV, and 2.1~GeV incident energies. The total
experimental resolution is
$\sigma \approx$ 0.9~MeV.\\
\hspace*{4.mm}The resolution of the missing mass was measured in the
neutron missing mass from the p~p~$\to$~p~$\pi^{+}$~X reaction at forward
angles. Figure 6 shows the missing mass spectrum, on a logarithmic scale,
when the incident
proton beam was T$_{p}$=1.805~GeV and the spectrometer angle
$\theta$=0.75$^{0}$. Here, the experimental resolution is 
$\sigma \approx$ 2.2~MeV.
The resolution of the missing mass lowers for increasing angles.\\
\hspace*{4.mm}We clearly see three structures in Fig. 6, between the
neutron peak and the $\Delta$ bump. They are unexpected since there is no
room for new low mass and narrow baryons within the many quark models
\cite{cap}. Many checks were performed to ascertain the real
presence of these new physical structures and they will be discussed in
detail in Sec. V.\\
\section{CROSS SECTION DETERMINATION}
Several normalizations were performed in order to extract
the cross sections and
several corrections were applied which take into account 
accidental coincidences, lost events,
angular vertical acceptances, and drift chamber and trigger efficiencies.
The beam flux was calibrated with the help of two telescopes in direct view of
the target and an ionisation chamber downstream of the beam.
These three detectors were calibrated
using $^{12}$C activation measurements.
The data must also be normalized to the two detected particle momenta
ranges. Indeed these momentum dependant acceptances vary for
different missing masses
(or invariant masses),
different reactions, spectrometer angles and different
incident energies. Figure 7 shows
these variations for the missing mass of the 
p~p~$\to$~p~$\pi^{+}$~X reaction at
T$_{p}$=1.805~GeV and $\theta$=0.75$^{0}$. This normalization was done
for all cross sections presented. The momentum dependant acceptances were
obtained from analytical relations or numerical fits to real or simulated
data (see later in section V.A).
We have checked that
the same final cross sections were obtained independently of the choice made
for this normalization.
The empty line in the figure will be discussed in the next paragraph.\\
\section{SIMULATION}
A simulation code was written in order to study the properties of the
spectrometer and detectors. All experimental parameters were found to be
well understood, and no narrow structures are observable in the simulated
histograms. The simulation allows us to evaluate the corrections for the 
lost events.
Vertical angular acceptance corrections were previously determined with GEANT.
The main events lost are those where both
 particles detected have the same momentum. These events define
a narrow empty valley that can be seen clearly in Fig. 7. Such a small region,
where important losses occur, is without
consequence when M$_{X}$ from p~p~$\to$~p~$\pi^{+}$~X is considered,
but important for a narrow region in M$_{p\pi^{+}}$ (mainly for neutron
missing masses). This region of
invariant mass was eliminated from further consideration in order to avoid
the introduction of fake
structures via a substantial renormalization.\\
\hspace*{4.mm}Figure 8 shows a comparison between the neutron
missing mass data and the corresponding simulation for
the p~p~$\to$~p~$\pi^{+}$~X
reaction at T$_{p}$=1.52~GeV and $\theta$=0$^{0}$. Inserts (a) and (b)
show data, whereas inserts (c) and (d) show the corresponding simulated
results. Inserts (a) and (c) show that these events came from the total
momentum range (0.6~$\le$pc$\le$~1.4~GeV). Both neutron missing
mass peaks are shown in inserts (b) and (d). They are regular and do not
display any internal  structures. Had there been dead wires in the
detectors, we would have expected regions of varying density. This is
not observed. There is no evidence either of the opposite situation of
self excited wire amplifiers. In Fig. 8 the cross section variation for
p~p~$\to$~p~$\pi^{+}$~n which differs between data and simulation,
explains the difference in intensity between inserts (a) and (c).\\ 
\section{CHECKS}
A careful internal calibration of all elements was undertaken,
and all possible cross checks were carried out in order to
evaluate the level of confidence that can be attributed to the existence of
the observed structures. In the following we provide the description of the
used procedures.\\
\subsection{Selection of different momenta ranges}
Several tests were performed in order to verify that the
structures were not produced
in a limited region of momenta. The data from the p~p~$\to$~p~$\pi^{+}$~X
reaction at T$_{p}$=1.52~GeV and forward angles were used. The neutron peak
was removed by software cuts in order to enhance the missing mass region of
interest. The statistics are consequently
reduced but the figures 9, 10, and 11 show that the
structures remain \cite{str}.\\ 
\subsection{Selection of different angular ranges}
Different angular cuts were performed during the analysis in order to test that
our structures remain
for different angular ranges inside our horizontal aperture of
$\pm$ 50~mrad. Figure 12  shows that is indeed the case. The data of the
p~p~$\to$~p~$\pi^{+}$~X
reaction at T$_{p}$=1.805~GeV and $\theta$=0.75$^{0}$ are presented in four
different angular ranges. Here the range is defined around the
mean horizontal direction of the spectrometer aperture.\\
\subsection{The structures do not depend on the spin state of the incident
beam} Figure 13 shows the missing mass of the
$\vec{\mbox{p}}~p~\to~p~\pi^{+}$~X
reaction at T$_{p}$=1.805~GeV and $\theta$=0.75$^{0}$. The inserts
(a) and (b) correspond to the two spin states of the incident particles.
We observe that the structures are present in both cases.\\
\subsection{Structures not present in accidental coincidences}
Figure 14 shows the comparison between the real events and the accidental
coincidences of the
missing mass from the  p~p~$\to$~p~$\pi^{+}$~X
reaction at T$_{p}$=1.805~GeV and $\theta$=0.75$^{0}$. The accidental
coincidences are defined
by software cuts in a range 12 times larger  than the true p-$\pi^{+}$
coincidences. The corresponding histogram does not display any structure.\\
\subsection{Empty target measurements}
The effect of the target windows was checked by regular
 empty target measurements performed during the data taking. The
count rates were always small, typically
 $\le$5\%. Figure 15  shows one comparison of full
target to empty target
measurements. The missing mass data from the p~p~$\to$~p~$\pi^{+}$~X
reaction at T$_{p}$=1.805~GeV and $\theta$=0.75$^{0}$ are presented.
Here both spectra
correspond to the same incident proton flux. We observe the absence of
structures in the empty target data. We deduced that the target windows were
not a source of a noticeable contamination. We also deduced that although
our measurements were performed at small angles, the data were not
contaminated by any hot area of incident beam which could have been scattered
by some mechanical element at the entrance of the spectrometer.\\
\subsection{Possible effect of particles emitted outside the
spectrometer solid angle and slowed down}
We consider here the possibilty to attribute the narrow structures to an
eventual slowing down of the detected particles (p and-or $\pi^{+}$).
The particles could be
emitted outside the solid angle and then could be
partially absorbed by the lead diaphragm
and-or other mechanical elements at the entrance of the spectrometer.
A careful simulation of this possible effect was performed. There is no
problem in the horizontal plane. But since the vertical angular
magnification of the spectrometer is small ($\approx$0.14),
the drift chambers were
not precise enough to make clear vertical cuts. The spectrometer aperture
is $\pm$~50~mrad. Trajectories
emitted vertically up to $\pm$~80~mrad were simulated
 and particles emitted with momenta up to 2~GeV/c were studied (the
maximum detection  momentum is 1.4~GeV/c). Figure 16 shows the result of
the simulation normalized to the missing
mass of the neutron peak (logarithmic scale). There is clearly no
problem outside the
1.06$\le M_{X} \le$1.11~GeV range for the missing mass of the
p~p~$\to$~p~$\pi^{+}$~X
reaction at T$_{p}$=2.1~GeV and $\theta$=0.7$^{0}$.
A similar result was found \cite{dub98} for a comparable simulation performed
under different experimental conditions, namely the p~p~$\to$~p~$\pi^{+}$~X
reaction at T$_{p}$=1.805~GeV and $\theta$=0.75$^{0}$. A broad peak is observed
inside the range 1.06$\le M_{X} \le$1.11~GeV
for eventual fake events slowed down by the lead diaphragm.\\
\hspace*{4.mm}Trajectories for increasing angles
($\left|\Delta~\phi\right|\ge$~50~mrad)
are progressively cut by the magnet, the yoke and the detectors. For the
detectors, this is particularly true at both extremities, since
the trajectories envelope fit exactly the trigger
dimensions, which correspond to $\Delta \phi~=~\pm$50~mrad. Figure 11
illustrates clearly the presence of structures in these extreme momenta
regions, where the fake effect
discussed here cannot be present.\\
\hspace*{4.mm}There are additionnal arguments excluding this effect
which could attribute the narrow structures to particles
slowed down by the slits. If the structures were produced by the neutron peak
shadows, then:\\
- for increasing spectrometer angles, the simulation shows that: \\
\hspace*{6.mm}- the mass of the broad fake peak will increase (up to
20~MeV),\\
\hspace*{6.mm}- its width will increase (up to 75~MeV);\\
- the ratio of the cross sections of the structures versus the neutron
missing mass
cross sections should be flat;\\
- the analyzing powers of the structures and of the neutron missing mass
peak should always be equal.\\
\hspace*{4.mm}None of these characteristics are observed. We concluded
therefore that this type of  contamination was not present.\\
\section{RESULTS}
All the checks described till now lead us to conclude that 
genuine structures are observed. Each corresponding mass and width
was obtained using a polynomial distribution for the background and a
gaussian one for the peak. The precision of the extracted peaks was
determined by the corresponding number of standard deviations (S.D.):
\begin{equation}
\hspace*{-1.cm}
S.D.=1/\sqrt{2} \sum_{i=1}^n [(N_{Ti}-N_{Bi})/ \Delta\sigma_{i}^2]~~/~~
[\sum_{i=1}^n(1/\Delta\sigma_{i}^2)]^1/2
\end{equation}

\hspace*{-4.mm}where the sum is done over the n different channels
describing the peak.
$N_T$ and $N_B$ describe the total and background count rates
of each channel,
$\Delta\sigma_{i}$ is the corresponding statistical error and
the factor 1/$\sqrt{2}$ comes from the assumption that the error
on $N_B$ is the same as the error on $N_T$.\\
\hspace*{4.mm}The next step in the analysis is to try to study to what 
they can be connected, and understand where these structures come from.
The main point will be to see if,
whether or not, they appear at stable masses.
The answer will be given after looking at the
different results.\\
\subsection{Ranges of baryonic mass studied by different reactions}
Figure 17 shows the useful range of baryonic missing masses studied 
below M=1.46~GeV. The three (two) narrow strips for each reaction
correspond to the three (two) different incident energies, increasing from
left to right, for the reactions studied. The variations
of these limits for different spectrometer
angles are small and therefore not pointed out. The statistics
 are not regularly distributed. At both sides of each range
the statistics are poor, therefore narrow structures could exist
in these cases, and have not been
extracted in this study. The short horizontal lines indicate the mass
of the observed narrow structures to be discussed in the following
paragraphs.\\
\hspace*{4.mm}It is sometimes possible to observe small
shoulders, or peaks, in publications of data obtained from
various reactions studied for
different aims. The corresponding
experimental resolutions are usually lower than in our experiment.
These shoulders or
peaks are generally not commented on by the authors. In a few cases they are
assumed (without calculation) to be the rescattering of particles
in the final state. The cross sections of the np$\to$pX reaction,
measured at $0^{0}$ at nine incident energies at LAMPF \cite{bon},
exhibit several structures. For example in this publication a peak can be
observed (at $T_{n}$=673~MeV, i.e. $M_{X}\approx$ 1136~MeV 
and $\sigma~\approx$~10~MeV).\\ 
\hspace*{4.mm}Some other results will be discussed later in sections VI.E,
F, and G, and their
data will be shown.\\
\subsection{The missing mass of the p~p~$\to$~p~$\pi^{+}$~X reaction}
The first results from this reaction were already published \cite{bor1},
\cite{dub98}. Figure 18 shows the scatterplot of proton momenta against pion
momenta from $T_{p}$=2.1~GeV and $\theta$=0.7$^{0}$ events. The same events,
after kinematical transformations, are shown in Fig. 19 where the
missing mass $M_{X}$ is plotted
versus the invariant $M_{p\pi^{+}}$ mass. In both figures, we
see clearly the neutron missing mass and the intense broad
$\Delta-\Delta$ region. Several narrow lines can be perceived
at fixed missing masses between the neutron and $\Delta$ missing
masses.
The empty narrow
band, as already explained in section IV, corresponds to lost events
where both the p and $\pi^{+}$ have the same momentum. Figures 20 and 21
 show the corresponding data at lower energy:
$T_{p}$=1.805~GeV and $\theta$=0.75$^{0}$. The data in Fig. 21, projected
onto the missing mass axis, are shown in Fig. 6. We see clearly the neutron
and the $\Delta$ peaks, and also small structures between the two. In
order to enhance these structures, we remove the neutron peak by applying 
software
cuts. Figure 22 shows the result for the p~p~$\to$~p~$\pi^{+}$~X reaction
at $T_{p}$=1.805~GeV and
$\theta$=0.75$^{0}$ after this selection, whereas Fig. 23 shows
the cross section at the same energy
but for $\theta$=3.7$^{0}$. A good definition of the structures
with respect to the background is obtained.\\
\hspace*{4.mm}For these structures to have any meaning we would expect them
to appear at the same masses, independently of the spectrometer angle and
independently of the incident energy.
In order to see if this is true, the cross sections as a function of missing
mass, are shown in figures 24 and 25 for different forward angles and at
all three proton beam energies.
 The different results are arbitrarily separated by a
translation to allow a better observation. In both figures the vertical
straight lines are drawn
at the same masses. We observe that the structures are present in nearly all
spectra, and when present, are observed at fixed masses: 1004~MeV,
1044~MeV, and 1094~MeV. The associated number of standard deviations (S.D.)
vary between 2 and 16.9 \cite{bor1}. Due to the large
background, the widths of the structures, and therefore their production
 cross sections, are inextricable.
 The mass of the first two structures is
M$\le$M$_{N}+M_{\pi^{+}}$. It is impossible to extract the widths of
these structures since they are
expected to be much smaller than that attainable by the experimental
resolution. These widths are narrow since no strong interaction
disintegrations can occur at these masses.\\ 
\hspace*{4.mm}When all three structures at M=1004~MeV, 1044~MeV, and 1094~MeV
are observed at nearly all angles and energies, narrow structures are also
observed at higher masses, but lightly excited and therefore more rare.
Consequently they may be observed with a small
 S.D. and so making them less certain. They are kept when observed several
 times at neighbouring mass values. These results are presented in
 the following subsections. The final mass is the mean value of the
 masses observed when the structures were present. The results will
 be presented later in section VII.A, regrouping 
 the different spectra, from different reactions, for each narrow structure
 mass.\\
\subsection{The invariant masses from the p~p~$\to$~p~$\pi^{+}$~X reaction.}
Only the results from the M$_{p\pi^{+}}$ invariant masses will be discussed.
The M$_{X\pi^{+}}$ invariant masses are larger than M=1.46~GeV and the
corresponding results will be presented in a forthcoming paper. The maximum
mass range of the M$_{p\pi^{+}}$ varies with the incident proton energy
(and is slightly dependent on the spectrometer angle) as indicated
in Fig. 17. This upper
limit is imprecise since the count rates at the upper edge of the
M$_{p\pi^{+}}$ invariant mass are very low. Again, in this mass range
several narrow structures were observed although less excited than
in M$_{X}~\le$~1.1~GeV. They were subsequently observed in certain
kinematical configurations and sometimes with a low S.D. and so are less
certain. These results will be discussed later (see later Fig. 36, Fig. 37,
and Fig.38).\\
\subsection{The missing mass from the $\vec{\mbox{d}}$~p~$\to$~p~p~X reaction.}
This reaction was studied some time ago \cite{bor2}. First results on the
tensor analyzing powers and vector analyzing powers of the invariant
M$_{pp}$ mass were studied with the aim of searching for narrow dibaryons.
Small structures in the missing mass M$_{X}$ were observed,
but their low S.D. prevent us from concluding that narrow baryons
were observed. The situation is different today since the masses of the
narrow structures once observed, are
very close to those observed now in the more recent
p~p~$\to$~p~$\pi^{+}$~X measurements.
Therefore, the results from the $\vec{\mbox{d}}$~p~$\to$~p~p~X reaction are 
consistent with the observed structures.\\
\hspace*{4.mm}These
results are shown in figures 26, 27, 28, 29, 30, 31, and 32.
They will be discussed in the next subsections. Figures 26 and 27
 display the number of events (as opposed to cross sections), since
 initially corrections for inefficiencies and acceptances were not
available. The migration to more modern computing systems has prevented
further analysis of this data since it became impossible to read
the data from the original magnetic tapes. The lack of any sharp variation
in the acceptance corrections, as now observed in the
p~p~$\to$~p~$\pi^{+}$~X data analysis, allows us to consider these
spectra.\\

\subsection{Invariant mass distributions of the
$\gamma~n\to~p~\pi^{-}\pi^{0}$ reaction.}
Different reactions of double-pion photoproduction on a nucleon were studied
experimentally and analyzed theoretically \cite{lag}. The multitude of
baryonic
resonances which must be considered in the intermediate states produce
different channels which can
interfere, and consequently an oscillatory cross section shape can be
observed. However, these oscillations are broad \cite{lag}
in comparison to the widths considered in the
present work and cannot be the cause of eventual
structures having $\sigma \approx$10~MeV. The data from such
measurements,
performed in order to study the reaction mechanism in double-pion
photoproduction, cannot generally be used for the present study because
of the large spacing between adjacent N$\pi$ invariant masses.
This is the case for the total cross sections of the following reactions:
$\gamma~p\to~p\pi^{-}\pi^{+}$, $\gamma~p\to~n\pi^{+}\pi^{0}$, and
$\gamma~p\to~p\pi^{0}\pi^{0}$ measured at MAMI \cite{bra}.
This is also the case for the Compton scattering by the proton
\cite{wol} measured at MAMI, and references therein.\\
\hspace*{4.mm}
An exception to this is the $\gamma~n\to~p~\pi^{-}\pi^{0}$ reaction.
This reaction was studied by Zabrodin {\it et al.} \cite{zab} at MAMI in the
photon energy range 500$\le~T_{\gamma}~\le$800~MeV. The
(p$\pi^{0}+p\pi^{-}$) spectra  were integrated over 50~MeV bins of the
incident photon
energy. The experiment was clearly not performed with the purpose of
looking for
eventual narrow baryons, and therefore the statistics for this purpose are
poor.
However, some structures can be observed which were not discussed
by the authors. With the aim of seeing
where they lie with respect to the other narrow structures, we extracted
these peak values.
Figure 33 shows a selection of the results of \cite{zab}, and table III
describes the range of integration of the incident photon energy, the mass
extracted and the corresponding S.D. Each peak will be discussed
separately in this text in subsections I to P.\\
\subsection{Cross section of the $\gamma$p$\to \pi^{+}$n reaction}
This reaction was studied at the Bonn 2.5~GeV electron synchrotron,
using a photon incident energy of 0.3$\le~T_{\gamma}\le$2.1~GeV \cite{dan}.
The experiment was performed in order to study the electromagnetic structure
of the nucleon resonances formed in the s-channel.
The excitation functions were measured at six different backward angles.
At all angles
the cross sections decrease quickly up to a total energy of W=1.3~GeV, and
some narrow structures can be observed at larger masses. The cross sections
were integrated over two channels, and fits of these narrow structures
in the range 1.3$\le W\le$ 1.44~GeV were performed. Figure~34 shows some of
the results obtained and correspond to four angles, whereas table IV
describes the
quantitative values extracted from the previous fits. The result
for each peak will
be discussed separately in this text in subsections I to P.\\
\subsection{The pd$\to$ppX reaction}
This reaction was studied with the 305~MeV proton beam of the Moscow Linear
Accelerator \cite{fil}. The authors observed narrow dibaryons in the pX
system and narrow baryonic peaks at 966~MeV, 986~MeV, and 1003~MeV
in the missing
mass M$_{X}$. They suggest that the narrow baryons observed at masses below
the pion disintegration threshold mass, ``are not excited states of the
nucleon, but resonancelike states caused by possible existence and decay of
narrow dibaryons. They cannot give contribution to the Compton scattering on
the nucleon''. Their largest mass: 1003 MeV, corresponds to our lightest
narrow exotic structure at M=1004 MeV. Their two lightest masses
correspond to a range not studied in the present work.\\
\subsection{Structure at M=1004~MeV}
A structure at this mass can be perceived with varying clarity in the
missing mass of the
p~p~$\to$~p~$\pi^{+}$~X reaction at the three energies (Figs. 22, 23, 24,
and 25). This low
mass is outside the range of the invariant M$_{p\pi^{+}}$ mass of
the p~p~$\to$~p~$\pi^{+}$~X and also outside the range of the missing
mass of the $\vec{\mbox{d}}$~p~$\to$~p~p~X reaction at $T_{d}$=1722 and 2100~MeV.\\
\hspace*{4.mm}Two shoulders were observed in the missing mass $M_{n}$
of the $p_{d}p_{i}\to\pi^{+}$pn reaction (deuteron vertex) in an
experiment performed at Dubna using 3.34~GeV/c deuterons \cite{hla}.
The masses of these
shoulders: M$\approx$ 1.0~GeV and M$\approx$ 1.05~GeV are
close to the masses of our two first narrow baryons. A peak at the same mass
was observed in the pd$\to$ppX reaction \cite{fil} (see previous subsection
VI G.).\\
\subsection{Structure at M=1044~MeV}
A small structure at this mass was observed in the missing
mass of the
p~p~$\to$~p~$\pi^{+}$~X reaction at all three incident energies studied
(Fig. 22, 23, 24, and 25). This low
mass is also outside the range of the invariant M$_{p\pi^{+}}$ mass of
the p~p~$\to$~p~$\pi^{+}$~X. There is an indication (with
a small S.D.) of the presence of a structure in the missing mass $M_{X}$
of the $\vec{\mbox{d}}$~p~$\to$~p~p~X reaction.
This can be seen in insert(a) of Fig. 26, and
in the tensor analyzing power (Fig. 29) and in the
vector analyzing power (Fig. 31).\\
\hspace*{4.mm}A shoulder at this mass was also observed in the missing
mass $M_{n}$ of the previously cited Dubna experiment, namely the
$p_{d}p_{i}\to\pi^{+}$pn reaction \cite{hla}.\\ 
\subsection{Structure at M=1094~MeV}
A structure at this mass is clearly observed in the missing
mass of the
p~p~$\to$~p~$\pi^{+}$~X reaction (figures 22, 23, 24, and 25). As an example,
let us consider some quantitative information concerning this
missing mass peak.
At T$_{p}$=1.805~GeV and $\theta$=9$^{0}$, we have $M_{X}$=1092.3,
$\sigma$=10.7~MeV,
and S.D.=8.1. This low
mass is also outside the useful range of the invariant M$_{p\pi^{+}}$
mass of
the p~p~$\to$~p~$\pi^{+}$~X since it is too close to threshold.
There are some small indications of the
presence of a different amplitude around this mass in several
analyzing powers
of the $\vec{\mbox{d}}$~p~$\to$~p~p~X reaction. They appear
in Fig. 28 and in inserts (c) and (d) of Fig. 29.\\
\hspace*{4.mm}Figure 33 shows two structures in this mass region,
extracted from the $\gamma$~n~$\to~\pi^{-}~\pi^{0}$~n reaction
studied at MAMI \cite{zab}. The invariant p$\pi$ masses found 
(1086.5~MeV and 1080.6~MeV and in table III ) are
too small and imprecise since they are close to threshold
($\approx$1075.5~MeV).
The $\gamma~p~\to~\pi^{+}~\pi^{0}$~n reaction was also studied at
MAMI \cite{lan}. Although the authors said that ``deviations of the
experimental data from the phase space distribution are evidence for
resonant or meson intermediate states in the $\pi^{+}~\pi^{0}$ production'',
it is useful to point out that in M$_{n\pi^{0}}$ and M$_{n\pi^{+}}$
invariant masses, a narrow peak at M$\approx$1.095~GeV is
observed at low incident photon energies (their Fig. 2). The virtual compton
scattering in the nucleon resonance region was studied at JLAB (Hall A) 
\cite{lav}.
A small peak ($\sigma~\approx$~12~MeV) was observed
(insert (a) of their Fig. 3) at W$\approx$1.098~GeV.
The pp$\to$pp$\pi^{0}$ reaction was studied at Uppsala \cite{bil}.
In this work the
p$\pi^{0}$ invariant mass displays a good agreement between measured and
Monte-Carlo simulations, except in the 1090-1100~MeV mass region (their
Fig. 18). So all these combined, independent observations point to the same
conclusion, namely an unidentified state at 1094 MeV.
\subsection{Structure at M=1136~MeV}
This was observed in the $M_{p\pi}$ invariant mass at $T_{p}$=1805~MeV
and 2100~MeV (see Table V). Figure 35 shows four
inserts where a small peak was extracted from the data in this mass region.
Table V indicates the masses found, the corresponding experimental widths,
the number of S.D., the incident energy, and the spectrometer angle.\\
\hspace*{4.mm}A small structure at M$_{X}$=1140~MeV was observed in the
$\vec{\mbox{d}}$~p$\to$p~p~~X reaction (insert (c) of Fig. 26).
Small structures in
this mass region were extracted at M$\approx$1129~MeV and 1146~MeV, from the
$\gamma$~n$\to$~p~$\pi^{-}\pi^{+}$ reaction (inserts (a) and (b) of Fig. 33,
and Table III). 
\subsection{Structure at M=1173~MeV}
Figure 36 shows four inserts showing a shoulder at this mass in the invariant
M$_{p\pi^{+}}$ spectrum. Table VI describes the corresponding results. This
structure is not observed in the missing mass. Only at T$_{p}$=1805~MeV and
$\theta=3.7^{0}$ is there a very small peak at 1162~MeV, but with a
very low S.D. (S.D.=1).\\
\hspace*{4.mm}A small structure at M$_{X}$=1178~MeV was observed in
$\vec{\mbox{d}}$~p$\to$p~p~~X reaction at $T_{d}$=1722~MeV (insert (d)
of Fig. 26). In
the same reaction at $T_{d}$=2100~MeV, a structure was extracted
at 1171~MeV (insert (c) of Fig. 26).\\
\hspace*{4.mm}A small peak was observed at 1184~MeV (insert (d) of Fig. 33),
 in the $\gamma$~p$\to$~p$\pi^{+}\pi^{-}$ reaction studied at
 MAMI \cite{zab}.\\
\hspace*{4.mm}There is an indication of a  small and narrow structure
at W$\approx$1168~MeV, in the total $\pi^{0}$ electro-production cross
section studied
at JLAB (Hall A) \cite{lav} (see $\sigma_{Tot}$ in their Fig. 1).
In the same paper there is also a small indication at the same mass
observed in the virtual Compton scattering excitation curve 
(at $\Phi=0^{0}$).\\
\subsection{Structure at M=1249~MeV}
Figure 37 is comprised of four inserts showing a shoulder in the invariant
M$_{p\pi^{+}}$
mass from the p~p~$\to$~p~$\pi^{+}$~X reaction at $T_{p}$=1.52~GeV.
Table VII describes these results.\\
\hspace*{4.mm}A shoulder was also observed in the missing mass of the
$\vec{\mbox{d}}$~p$\to$p~p~X reaction (insert (a) of Fig. 27 and Table II).\\
\hspace*{4.mm}There is a very small indication of a structure at a mass
close to 1246~MeV in the total $\pi^{0}$ electro-production cross section
studied at JLAB (Hall A) \cite{lav}.
\subsection{Structure at M=1277~MeV}
Figure 38 shows three inserts where a shoulder in the invariant
M$_{p\pi^{+}}$ mass and in the missing mass of the p~p~$\to$~p~$\pi^{+}$~X
reaction is observed. Table VIII regroups these results.\\
\hspace*{4.mm}A shoulder was extracted at 1272~MeV in the
$\gamma$~p$\to$~p$\pi^{+}\pi^{-}$ reaction studied at MAMI \cite{zab}.
Figure 33 (insert (c)) and Table III show the associated values.\\
\subsection{Structure at M=1339~MeV}
A small structure in the missing mass $M_{X}$=1327.1~MeV
was observed at $T_{p}$=2.1~GeV and $\theta=9^{0}$ (Fig. 38 (d)). This mass
should normally not be mentioned from this single result, but peaks at nearby
masses were observed in previously
published data from other experiments. Two structures
were extracted from the np$\to$pX cross section \cite{dan} at M=1337~MeV
 for $\theta=120^{0}$ and at 1345~MeV for $\theta=135^{0}$ (see Fig. 34).
 A peak was extracted from the $\gamma$p~$\to~\pi^{+}$n
reaction \cite{zab} at M=1347~MeV (see Fig. 34).\\  
\subsection{Structure at M=1384~MeV}
Figure 39 shows two inserts where a shoulder in the missing mass of the
p~p~$\to$~p~$\pi^{+}$~X and d~p~$\to$~p~p~X
reactions is observed. Table IX describes the results.
A structure at M=1369~MeV, $\sigma$=14.5~MeV, and S.D.=3.6 was 
already observed in Fig. 38 (d).\\
\hspace*{4.mm}A shoulder was extracted at 1392~MeV from the
$\vec{\mbox{d}}$~p$\to$p~p~X reaction (insert (b) of Fig. 27 and Table II).\\
\hspace*{4.mm}A peak at the same mass was extracted from the 
cross sections of the $\gamma$p$\to \pi^{+}$n reaction studied at Bonn
\cite{dan}. It was observed at four angles. Table IV and Fig. 34 show the
results extracted after an
integration over two channels. The mean mass value from that experiment is
M=1385.8~MeV, very close to the mass 1384~MeV extracted from our data.\\
\hspace*{4.mm}In the chiral constituent quark model, the calculated
spectrum of N-like $qqqq\bar{q}$ states \cite{hel} exhibits 
 two states $J^{P}$=1/2+ and 3/2+,
 (T=1/2) at $M\approx$1366~MeV. This mass is not very far
from our experimentally observed state at 1384~MeV. There is no indication
concerning the width of this calculated state. Since there are no lower
calculated masses, an identification between this calculated state and the
one found experimentally at M=1384~MeV, is rather improbable.

\section{DISCUSSION}
\subsection{General discussion}
Three strongly excited structures were observed in the missing mass
spectra at
$M_{X}$=1004~MeV, 1044~MeV, and 1094~MeV. Several lightly excited
structures were
observed at heavier masses. They were considered when observed several times
at the same mass to within a few MeV. Their non-observation in other spectra
is associated with their small cross section, making them difficult to
extract or observe. The existence of these structures is
strengthened by the fact that they are usually observed in
invariant masses. Indeed the confidence on small peaks observed in missing
masses may be smaller. Moreover these structures are observed in
various independent
experiments.  Some peaks were not considered since they were not observed
several times in our data. A good example of such a case is the spectrum of
$M_{p\pi}$
invariant mass at $T_{p}$=1.805~GeV and $\theta=0.75^{0}$, where a
structure was extracted at 1232~MeV.\\
\hspace*{4.mm}Figure 40 shows all masses discussed previously along with
those presented from our
measurements as well as those extracted from previously published data 
from different
experiments. The horizontal bands correspond to the mean mass $\pm$3~MeV.
Table X allows to connect the different experiments to the marker plot on
Fig. 40.
It is worth noting that very few mass structures lie outside the
range defined by the horizontal bands of Fig. 40.\\
\hspace*{4.mm}The only high precision experiment previously dedicated to
looking for narrow baryons was
performed at TRIUMF \cite{ram}. The reaction studied was p~p~$\to~B^{++}$~n
(in fact p + CH$_{2}~\to~B^{++}$~X) at 460~MeV.
In this experiment a hypothetical doubly charged baryon
$B^{++}$ was looked for behind a
spectrometer and was not observed in the limit of 0.75~pb/sr. There were two
reasons to look for a doubly charged baryon:\\
- if it exists, it would be easier to identify among a large flux of singly
charged protons and pions,\\
- since only a weak decay would be possible ($B^{++}~\to~p~e^{+}~\nu_{e}$)
its lifetime should be long enough, at least $10^{-2}$~s on the basis of the
hyperon lifetime, making it possible to detect behind a spectrometer
\cite{ram}. Since this hypothetical doubly charged baryon was not observed,
we concluded on the isospin 1/2 for our first 
exotic baryons observed. Such an isospin attribution agrees with our quark
cluster mass formula (see later, section VII.B) and with the diquark
model of Konno \cite{kon} (see later, section VII.C).\\
\hspace*{4.mm}Another experiment was studied before \cite{nak} at KEK, using
a 12~GeV proton beam, with the aim of
 looking for long lived exotic hadrons with charge $\pm$2 or -5/3. Once again
 no candidate was found. This experiment was performed
in somewhat unfavorable experimental conditions:
composite target (platinium), very long detection line (36 m), rather small
solid angle ($\Delta\Omega\approx$1 msr), and a small momentum dependent
acceptance ($\Delta$p/p$\approx \pm 3\%$). The negative result again agrees
with the isospin attribution (T=1/2) of our narrow, low mass exotic baryons.
Below M=1.2 GeV, only one calculated mass (at 1139~MeV) can have T=1/2 or 3/2
(see next subsection).\\
\hspace*{4.mm}Figure 17 shows the masses of the narrow structures observed in
this work. We note that there are only a few cases where a
structure could be observed and was not.\\
\hspace*{4.mm}Since we have not observed the structures in excitation
spectra with slowly increasing incident energy, we are sheltered from any
possible cusp effect (threshold energy for the production of any heavier
meson). Moreover the threshold energy for these cusp effects would be
different for different reactions, such
as p~p~$\rightarrow$~p~${\pi^+}X$ and d~p~$\rightarrow$~p~p~X. This effect
is the only one sometimes advocated to explain the experimentally observed
narrow structures.\\
\hspace*{4.mm}As already discussed, there is no
room for new, low mass and narrow baryons within the many quark models
\cite{cap}. These models use several parameters, and the baryonic
  masses calculated depend on the set of parameters chosen (see for example
\cite{gar}). Within the mass range discussed in this work, the number of
baryons found is equal to the number of experimental baryons for each
T and $J^{P}$ even if the calculated masses are sometimes much lower.
For example, in the calculation of \cite{gar}, the lowest
N~$1/2^{-}$ and N~$3/2^{-}$ states are found at masses close to 1360~MeV when
their first set of parameters is used.\\
\hspace*{4.mm}The experimental determination of the quantum numbers of 
the observed exotic baryons is outside the scope of the present work. 
This requires the knowledge of the
angular distributions. Also knowing the parity would be very
useful, since it would allow the verification of the assumption concerning
the number of
quarks (anti-quarks) involved.\\
\subsection{Quark cluster mass formula}
In our previous papers, dedicated to narrow dibaryon \cite{bor} and meson
\cite{bor4} searches,
it was shown that the masses of the experimentally observed narrow
structures were in good agreement with
those obtained through a simple phenomenological mass formula
using two-cluster $q^{n}\bar{q}^{m}-q^{p}\bar{q}^{r}$
configurations: 
\begin{equation}
\hspace*{-1.cm}
M = M_0+M_1[i_1(i_1+1)+i_2(i_2+1)+(1/3)s_1(s_1+1)+
(1/3)s_2(s_2+1)]
\end{equation}
where s$_{1}$(s$_{2}$) and i$_{1}$(i$_{2}$) are the first (second)
cluster spin and isospin values. This formula was derived some years ago
\cite{mul} for two clusters of quarks at the ends of a stretched bag in
terms of color magnetic interactions.\\
\hspace*{.4cm}The same approach is employed here. Equation (1) involves a
large degeneracy.
We made the assumption that the simplest configuration is
preferred, otherwise the possible spin and isospin will increase and the
parity will be degenerate since additionnal $q\bar{q}$ configurations will
always be possible. The two parameters, described in \cite{bor1}, were adjusted
in order to find the mass, spin, and isospin values of the nucleon and the
Roper N$^{*}$(1440) baryons. Such assumptions lead to the values
$M_{0}$=838.2~MeV and $M_{1}$=100.3~MeV. This does not mean that we consider
our new exotic baryons to be excited states of the nucleon. Indeed there
is no simple overlap
between $q^{3}$ and $q\bar{q}~-~q^{3}$, or $q^{2}~-~q$ wave functions.\\
\hspace*{.4cm}Figure 41 shows the masses, spins, and isospins 
obtained using equation (1). We observe the surprisingly 
good agreement
between the experimental masses and the masses calculated without any
adjustable parameter, for the four first states.
The agreement is in fact excellent for all masses, except two states
at M=1173~MeV and
M=1384~MeV. The absence of several masses, predicted by the previous
formula, can be related to the fact that they are weakly excited. It is not
excluded that more precise experiments will, in the future, observe these
states. Below the pion emission threshold at 1075~MeV, the only possible decay
channel is the radiative one, and the observed width is due to the 
experimental resolution.\\
\subsection{The diquark cluster model}
The diquark cluster model was developed over many years \cite{kon}.
It was used to calculate the masses of the narrow mesonic resonances,
the masses of narrow dibaryons, the mass of the exotic I=2 meson, and the
masses of narrow baryonic states. The model does not assume the existence of
colored quark clusters. Different parameters were used which were all
determined previously using the data of baryon masses and the ETH group
$\pi$d phase shifts. The model assumes that these baryons consist of
a diquark and a quark, and that the residual interaction between them is
negligibly weak. Figure 42 shows the masses of the exotic baryons 
calculated within this model, the isospin and spin of the levels, and the
comparison with the experimental masses observed in this work. 
With the exception of the levels at M=1173 MeV and M=1249~MeV, we see that
the masses calculated are close to the masses observed experimentally.
Once again, there are levels predicted but not observed, and it is
not excluded
that they exist and could be measured in dedicated experiments in the future.\\
\subsection{The metastable levels model}
It was assumed \cite{kob} that the first narrow baryons (with masses lower
than 1075~MeV), are metastable
levels in the three quark system, as a member of a total antisymmetric
representation of the spin-flavor group
($\underline{20}$-plet of the {\it SU}(6)$_{FS}$) .
Since such states
cannot be excited by one photon or decay to $\gamma$N, the simplest
decay channel is assumed to be 2$\gamma$N. In that case, they will not
contribute to Compton scattering.\\
\subsection{Excitation of collective states of the quark condensate}
A model was proposed by T. Walcher, which associates the narrow baryonic
states below the $\pi$ threshold production to multiproduction of a
genuine virtual
Goldstone Boson with a mass close to 20 MeV \cite{wal}. This model explains
not only the combined experimental results for the narrow nucleon states
presented above, but also the states observed in the missing mass $M_{X}$
of the pd~$\to$~ppX reaction \cite{fil} and the rather equidistant level
spacing of the narrow dibaryons observed experimentally \cite{bor}.\\
\subsection{The chiral-scale bag model}
This (CSB) model \cite{pok} deals with chiral and scale symmetries of QCD and
their violations. It predicts small radii color confinement, and found that
the ground states of {\it SU}(3) hadrons could be treated as
predominantly multi~bag states ({\it Bag$\overline{Bag}$Bag} for baryons).
This picture allows low mass, narrow excitations with a mean distance
$\approx$50~MeV between the almost degenerated spin-isospin states
that appear in strongly bound systems dealing with {\it Bag}
and {\it  $\overline{Bag}$}. These
narrow excitations are expected for all stable or narrow {\it SU}(3) hadrons.
\section{CONCLUSION}
Narrow baryonic structures were observed in the missing mass and
in the invariant mass of the p~p~$\to$~p~$\pi^{+}$~X and
d~p~$\to$~p~p~X reactions.
 All final data
(missing masses or invariant masses) were the results of two-particle detection.
Consequently, it is deduced that any inefficiency or hot point in any detector
is unable to produce a narrow peak.
If we leave aside the states with masses 1.0$\le$ M $\le$1.1~GeV,
the ratio of peak to
background is small. However, there is no structure without at least one
peak extraction with a number of S.D.~$\ge$~3.1 (see tables I to IX).
Many checks were undertaken which allowed to conclude that these structures
are genuine and not produced by 
experimental artifacts. Their masses and widths were extracted
using polynomials for the background and gaussians for peaks. The widths
found are not of a high precision since these peaks are usually much smaller
than
the physical background. The production cross sections are therefore not
precise, and are not discussed further.\\
\hspace*{4.mm}
The possibility of associating these peaks with dynamical
rescattering among final particles, was eliminated for two reasons:\\
-the structures are narrow,\\
-they are not spread but they are observed at stable masses
for different scattering angles and different incident energies.\\
\hspace*{4.mm}We concluded that these peaks correspond to new baryons.
There is no room for them within the many theoretical constituent quark
models. Indeed the chiral constituent quark model is known \cite{sim}
to yield a spectrum for the $qqq$ states which agrees well with the empirical
baryon spectrum up to M$\approx$1700~MeV. Moreover, there is no reason for
calculations, performed within the $qqq$ assumption, to explain the relatively
narrow widths observed in our data. Indeed, a calculation performed within a
chiral quark model, investigated the hadronic $\pi$ and $\eta$ decay modes
of N and $\Delta$ resonances \cite{the}. The authors found various decay
widths which were dependent on several calculation assumptions. However,
for the only baryon that they considered in the mass range of the present
study, the Roper resonance
N$_{1440}$ 1/2+, they calculated a width of the order of several
hundreds of MeV for the $\Gamma$(N$^{*}\to$N$\pi$) decay mode.\\   
\hspace*{4.mm}Therefore, we tentatively associate
our narrow baryons with exotic baryons made of two colored-quark clusters.
The experimental masses agree well with
the masses calculated within a phenomenological
relation derived twenty years ago for two colored clusters in
a spherical MIT
type bag. The formula is used as a phenomenological one, but allows to
determine the two parameters leading to the masses, spins, and isospin
of the nucleon and the Roper resonance. This does not mean that we consider
them to be excited states of the nucleon. Indeed, there is no simple overlap
between $q^{3}$ and $q\bar{q}~-~q^{3}$, or $q^{2}~-~q$ wave functions.\\
\hspace*{4.mm}Even if the hadronic picture dominated by the baryonic
resonances is the relevant degree of freedom, quarks are also relevant at
energies as low as the ones considered in this work, and at low momenta
transfer. Therefore, the belief of well separated regions, one for the 
quark-gluon picture and the other for the hadronic picture appears to be
too simple.\\

\acknowledgments
We are grateful to Dr. M. MacCormick for stimulating comments and
 help in writing our manuscript in English. We thank Dr. P. Pedroni 
who provided to us
the numerical values of the cross sections from the 
$\gamma$n$\to$p$\pi^{-}\pi^{0}$ reaction studied at MAMI.\\

\newpage
\begin{figure}
\caption{The SPES3 spectrometer and the associated detection system.}
\label{fig1}
\end{figure}
\begin{figure}
\caption{The MIT drift chamber efficiency. The triangles correspond to
the one-particle detection efficiency, whereas the circles correspond to the
double hit detection efficiency.}
\label{fig2}
\end{figure}
\begin{figure}
\caption{Time of flight spectra for four $A_{i}B_{j}$ combinations
consisting of the
number of events versus the channel number (125 ps/channel).
The four inserts (a), (b), (c), and (d) 
correspond respectively to the following mean momenta: 1.36~GeV/c, 1.25~GeV/c,
1.00~GeV/c, and 0.83~GeV/c.}
\label{fig3}
\end{figure}
\begin{figure}
\caption{Spectra of the second time of flight comprised of the number
of events versus the channel number 
(250 ps/channel) for all 190 $A_{i}A_{j}$ combinations (see text) of
the p~p~$\to$~p~p~X reaction at $T_{p}$=2.1~GeV and $\theta$=3$^{0}$.}
\label{fig4}
\end{figure}
\begin{figure}
\caption{Invariant mass of the p~p FSI peaks from the p~p~$\to$~p~p~X
reaction. Full circles, squares, and triangles correspond
respectively to data
(vertically shifted) obtained at T$_{p}$=1.52~GeV, 1.805~GeV, and 2.1~GeV.}
\label{fig5}
\end{figure}
\begin{figure}
\caption{Missing mass spectrum for the p~p~$\to$~p~$\pi^{+}$~X reaction at
T$_{p}$=1.805~GeV and $\theta$=0.75$^{0}$.}
\label{fig6}
\end{figure}
\begin{figure}
\caption{Scatter plot of the missing mass versus the proton momenta (insert
(a)), and the missing mass versus the pion momenta (insert (b)) from the
p~p~$\to$~p~$\pi^{+}$~X reaction at
T$_{p}$=1.805~GeV and $\theta$=0.75$^{0}$.}
\label{fig7}
\end{figure}
\begin{figure}
\caption{Neutron missing mass peak from the p~p~$\to$~p~$\pi^{+}$~X reaction
at T$_{p}$=1.52~GeV, $\theta$=0$^{0}$. Inserts (a) and (b) correspond to
data, inserts (c) and (d) are from simulation.}
\label{fig8}
\end{figure}
\begin{figure}
\caption{The p~p~$\to$~p~$\pi^{+}$~X reaction
at T$_{p}$=1.52~GeV and $\theta$=0$^{0}$. Selection of several momenta ranges
in inserts (a), (b), (c), and (d) respectively: p$_{\pi}~\ge$~1~GeV/c,
p$_{\pi}~\le$~1~GeV/c, p$_{p}~\ge$~1~GeV/c, and p$_{p}~\le$~1~GeV/c.}
\label{fig9}
\end{figure}
\begin{figure}
\caption{Missing mass of the  p~p~$\to$~p~$\pi^{+}$~X reaction
at T$_{p}$=1.52~GeV and $\theta$=2$^{0}$. The four inserts (a), (b), (c), and
(d) correspond to events selected by software analysis cuts in the following
momenta ranges 0.6~$\le~pc\le$~1.4~GeV (total range),
0.8~$\le~pc\le$~1.4~GeV, 0.6~$\le~pc\le$~1.2~GeV, and
0.8~$\le~pc\le$~1.2~GeV respectively.}  
\label{fig10}
\end{figure}
\begin{figure}
\caption{Missing mass of the  p~p~$\to$~p~$\pi^{+}$~X reaction
at T$_{p}$=1.52~GeV. In both inserts (a) and (b), a selection on detected
particle momenta is made in order to keep only
momenta where: p$\le$ 0.8~GeV/c or p$\ge$ 1.2~GeV/c. Both inserts
(a) and (b) correspond to $\theta$=2$^{0}$ and
$\theta$=5$^{0}$ measurements respectively.}
\label{fig11}
\end{figure}
\begin{figure}
\caption{Missing mass of the  p~p~$\to$~p~$\pi^{+}$~X reaction
at T$_{p}$=1.805~GeV, $\theta$=0.75$^{0}$.  The four inserts (a), (b), (c),
and (d) correspond to events selected by software cuts in the following
angular ranges respectively:
$\theta_{p} \ge$~0$^{0}$ and $\theta_{\pi^{+}} \ge$~0$^{0}$,
$\theta_{p} \le$~0$^{0}$ and $\theta_{\pi^{+}} \le$~0$^{0}$,
$\theta_{p} \ge$~0$^{0}$ and $\theta_{\pi^{+}} \le$~0$^{0}$, and
$\theta_{p} \le$~0$^{0}$ and $\theta_{\pi^{+}} \ge$~0$^{0}$.
Here the range is defined around the mean horizontal direction of the
spectrometer aperture.}
\label{fig12}
\end{figure}
\begin{figure}
\caption{Missing mass of the  p~p~$\to$~p~$\pi^{+}$~X reaction
at T$_{p}$=1.805~GeV and $\theta$=0.75$^{0}$. Inserts (a) and (b) correspond to
both spin states of the incident proton beam.}
\label{fig13}
\end{figure}
\begin{figure}
\caption{Missing mass of the  p~p~$\to$~p~$\pi^{+}$~X reaction
at T$_{p}$=1.805~GeV and $\theta$=0.75$^{0}$. True coincidences (full circles)
are compared to random coincidences (empty squares) defined over
a range 12 times greater.}
\label{fig14}
\end{figure}
\begin{figure}
\caption{Missing mass of the  p~p~$\to$~p~$\pi^{+}$~X reaction
at T$_{p}$=1.805~GeV and $\theta$=0.75$^{0}$. Comparison of full (squares)
and empty (circles)
target data, normalized to the same incident proton beam flux.}
\label{fig15}
\end{figure}
\begin{figure}
\caption{Effect of a possible slowing down of p and $\pi^{+}$ through
lead slits and stainless steel rings at the entrance of the spectrometer.
p~p~$\to$~p~$\pi^{+}$~X reaction at T$_{p}$=2.1~GeV and $\theta$=0.7$^{0}$.
Squares correspond to data whereas circles correspond to simulated 
slower particles, normalized by the same neutron missing mass peak.}
\label{fig16}
\end{figure}
\begin{figure}
\caption{Range of baryonic masses studied in different reactions.
The narrow strips correspond to the different incident energies
(increasing from left to right for each observed variable
and each reaction). The full
horizontal lines indicate the masses of the observed narrow structures.}
\label{fig17}
\end{figure}
\begin{figure}
\caption{Scatterplot of $p_{p}$ versus $p_{\pi^{+}}$ events from the
 p~p~$\to$~p~$\pi^{+}$~X reaction at T$_{p}$=2.1~GeV and $\theta$=0.7$^{0}$.}
\label{fig18}
\end{figure}
\begin{figure}
\caption{Scatterplot of missing mass $M_{X}$ versus invariant mass
$M_{p\pi{+}}$ from the p~p~$\to$~p~$\pi^{+}$~X reaction at T$_{p}$=2.1~GeV
and $\theta$=0.7$^{0}$.}
\label{fig19}
\end{figure}
\begin{figure}
\caption{Scatterplot of $p_{p}$ versus $p_{\pi^{+}}$ from the
 p~p~$\to$~p~$\pi^{+}$~X reaction at T$_{p}$=1.805~GeV and
 $\theta$=0.75$^{0}$.}
\label{fig20}
\end{figure}
\begin{figure}
\caption{Scatterplot of missing mass $M_{X}$ versus invariant mass
$M_{p\pi{+}}$ from the p~p~$\to$~p~$\pi^{+}$~X reaction at
T$_{p}$=1.805~GeV and $\theta$=0.75$^{0}$.}
\label{fig21}
\end{figure}
\begin{figure}
\caption{Cross section of the missing mass of the  p~p~$\to$~p~$\pi^{+}$~X
reaction at T$_{p}$=1.805~GeV and $\theta$=0.75$^{0}$.}
\label{fig22}
\end{figure}
\begin{figure}
\caption{Cross section of the missing mass of the  p~p~$\to$~p~$\pi^{+}$~X
reaction at T$_{p}$=1.805~GeV and $\theta$=3.7$^{0}$.}
\label{fig23}
\end{figure}
\begin{figure}
\caption{Missing mass spectra of the p~p~$\to$~p~$\pi^{+}$~X reaction
at T$_{p}$=1520~MeV and 1805~MeV and at the three smallest angles for both
energies.}
\label{fig24}
\end{figure}
\begin{figure}
\caption{Missing mass spectra of the p~p~$\to$~p~$\pi^{+}$~X reaction
at T$_{p}$=2100~MeV at the three measured angles.
From top to bottom: $\theta$=0.7$^{0}$, 3$^{0}$, and 9$^{0}$ lab.}
\label{fig25}
\end{figure}
\begin{figure}
\caption{Number of events in the missing mass spectra from the
$\vec{\mbox{d}}$~p~$\to$~p~p~X reaction.  The experimental conditions
corresponding to the inserts (a), (b), (c), and (d) are 
specified in Table I.}
\label{fig26}
\end{figure}
\begin{figure}
\caption{The $\vec{\mbox{d}}$~p~$\to$~p~p~X reaction at T$_{d}$=2.1~GeV and
$\theta$=17$^{0}$. The experimental conditions
corresponding to the inserts (a) and (b) are
specified in Table II.} 
\label{fig27}
\end{figure}
\begin{figure}
\caption{Tensor analyzing power of the missing mass of the
$\vec{\mbox{d}}$~p~$\to$~p~p~X reaction
at T$_{d}$=2.1~GeV and $\theta$=17$^{0}$.}
\label{fig28}
\end{figure}
\begin{figure}
\caption{Tensor analyzing power of the $\vec{\mbox{d}}$~p~$\to$~p~p~X reaction
at T$_{d}$=2.1~GeV and $\theta$=17$^{0}$. The four inserts (a), (b), (c), and
(d) correspond respectively to the following cuts applied on M$_{pp}$:
1876$\le~M_{pp}\le$1880~MeV ($^{1}S_{0}$) pp state,
1880$\le~M_{pp}\le$1883~MeV,
1883$\le~M_{pp}\le$1886~MeV, and 1886$\le~M_{pp}\le$1889~MeV.}
\label{fig29}
\end{figure}
\begin{figure}
\caption{Vector analyzing power of the $\vec{\mbox{d}}$~p~$\to$~p~p~X reaction
at T$_{d}$=2.1~GeV and $\theta$=17$^{0}$.}
\label{fig30}
\end{figure}
\begin{figure}
\caption{Vector analyzing power of the $\vec{\mbox{d}}$~p~$\to$~p~p~X reaction
at T$_{d}$=2.1~GeV and $\theta$=17$^{0}$. The inserts
have the same definition as those in Fig. 29.}
\label{fig31}
\end{figure}
\begin{figure}
\caption{Tensor and vector analyzing powers of the missing mass
of the $\vec{\mbox{d}}$~p~$\to$~p~p~X reaction
at T$_{d}$=1.722~GeV and $\theta$=0$^{0}$.}
\label{fig32}
\end{figure}
\begin{figure}
\caption{Invariant mass distributions of the
$\gamma$~n$\to$~p~$\pi^{-}\pi^{0}$ reaction [17] showing a selection of
cross sections where small narrow structures were extracted. The
experimental conditions corresponding to the inserts (a), (b), (c),
and (d) are specified in Table III.}
\label{fig33}
\end{figure}
\begin{figure}
\caption{Excitation functions of $\gamma$~p~$\to~\pi^{+}$~n cross sections
measured at Bonn [18], after a two-channel integration. The
experimental conditions corresponding to the inserts (a), (b), (c),
and (d) are described in Table IV.}
\label{fig34}
\end{figure}
\begin{figure}
\caption{A selection of cross sections where a narrow structure at
M$_{X}$=1136~MeV was observed. The experimental conditions
corresponding to the inserts (a), (b), (c), and (d) are 
described in Table V.}
\label{fig35}
\end{figure}
\begin{figure}
\caption{A selection of cross sections where a narrow structure at
M$_{p\pi}$=1173~MeV was observed. The experimental conditions
corresponding to the inserts (a), (b), (c), and (d) are
described in Table VI.}
\label{fig36}
\end{figure}
\begin{figure}
\caption{A selection of cross sections where a narrow structure at
M$_{X}$=1249~MeV was observed. The experimental conditions
corresponding to the inserts (a), (b), (c), and (d) are
described in Table VII.}
\label{fig37}
\end{figure}
\begin{figure}
\caption{A selection of cross sections where a narrow structure at
M$_{X}$=1277~MeV was observed. The experimental conditions
corresponding to the inserts (a), (b), and (c), are
described in Table VIII.}
\label{fig38}
\end{figure}
\begin{figure}
\caption{A selection of cross sections where a narrow structure at
M$_{X}$=1384~MeV was observed. The experimental conditions
corresponding to the inserts (a) and (b) are
described in Table IX.}
\label{fig39}
\end{figure}
\begin{figure}
\caption{Narrow structure baryonic masses observed in cross
sections from different reactions. These reactions are described
in Table X.}
\label{fig40}
\end{figure}
\begin{figure}
\caption{Narrow baryonic experimental and calculated masses.
Equation (1) was used (see text for more details).}
\label{fig41}
\end{figure}
\begin{figure}
\caption{Narrow baryonic experimental and calculated masses.
The calculation corresponds to the diquark cluster model [27].} 
\label{fig42}
\end{figure}
\newpage
\begin{table}
\caption{A selection of five missing mass cross sections from
the $\vec{\mbox{d}}$~p~$\to$~p~p~X reaction, showing narrow structures.
The columns describe each insert, the mass
found, the experimental width, the number of standard deviations of the
structure, the incident energy, the spectrometer angle, and the cuts
performed in the software analysis.}
\label{tab1}
\end{table}
\begin{table}
\begin{tabular} [h]{c c c c c c c}
Figure 26& Mass& width& S.D.&T$_{p}$&$\theta$&cuts on $M_{pp}$\\
 &(MeV)&(MeV)& &(GeV)&(MeV)\\
\hline
 (a)&1041&8.0&3.05&2.1&17$^{0}$&1876$\le M_{pp}\le$ 1889\\
 (b)&1094.7&17.4&3.0&2.1&17$^{0}$&1876$\le M_{pp}\le$ 1880\\
 (c)&1140.3&11.5&3.0&2.1&17$^{0}$&1876$\le M_{pp}\le$ 1880\\
 (c)&1171.2&10.3&3.95&2.1&17$^{0}$&1876$\le M_{pp}\le$ 1880\\
 (d)&1178.1&1.5&1.85&1.72&0$^{0}$&no cuts on $M_{pp}$\\
\end{tabular}
\end{table}
\begin{table}
\caption{A selection of two missing mass cross sections from the
$\vec{\mbox{d}}$~p~$\to$~p~p~X reaction, showing narrow structures.
The columns describe each insert, the mass
found, the experimental width, the number of standard deviations of this
structure, the incident energy, the spectrometer angle, and the cuts
performed with the software.}
\label{tab2}
\end{table}
\begin{table}
\begin{tabular} [h]{c c c c c c c}
Figure 27& Mass& width& S.D.&T$_{p}$&$\theta$&cuts on $M_{pp}$\\
 &(MeV)&(MeV)& &(GeV)&(MeV)\\
\hline
 (a)&1243.8&6.0&3.2&2.1&17$^{0}$&1876$\le M_{pp}\le$ 1880\\
 (b)&1391.9&2.8&2.6&2.1&17$^{0}$&1876$\le M_{pp}\le$ 1880\\
\end{tabular}
\end{table}
\newpage
\begin{table}
\caption{The
$\gamma$~n$\to$~p~$\pi^{-}\pi^{0}$ reaction studied at MAMI [18] (see section
VI.E).
 Cross sections of
the invariant p$\pi^{0}$+p$\pi^{-}$ mass. A selection of cross sections where
small narrow structures were extracted are presented and related to
the four inserts (a), (b), (c), and (d) of Fig. 33.}
\label{tab3}
\end{table}
\begin{table}
\begin{tabular} [h]{c c c c c}
Figure 33& Mass& width& S.D.&Energy range\\
 &(MeV)&(MeV)& &(GeV)\\
\hline
 (a)&1086.5&6&3.1&500$\le~T_{\gamma}\le$600\\
 (a)&1128.8&4.5&3.&500$\le~T_{\gamma}\le$600 \\
 (b)&1080.6&2.5&2.5&700$\le~T_{\gamma}\le$750\\
 (b)&1145.9&4.0&1.6&700$\le~T_{\gamma}\le$750 \\
 (c)&1272.3&6.6&2.9&700$\le~T_{\gamma}\le$750\\
 (c)&1347.2&4.7&1.9&700$\le~T_{\gamma}\le$750\\
 (d)&1184&2.7&2.9&500$\le~T_{\gamma}\le$800\\
\end{tabular}
\end{table}
\begin{table}
\caption{Excitation function of the $\gamma$~p~$\to~\pi^{+}$~n reaction
studied at the Bonn electron synchrotron [19].}
\label{table4}
\end{table}
\begin{table}
\begin{tabular} [h]{c c c c}
$\theta$~(degrees)&Mass (MeV)&width (MeV)&S.D.\\
\hline
95&1390.3&25.4&12.4\\
120&1336.9&8.7&6.5\\
120&1383.1&24.3&15.3\\
135&1345.0&11.3&7.7\\
135&1383.2&17.5&12.2\\
180&1386.6&19.1&4.7\\
\end{tabular}
\end{table}
\newpage
\begin{table}
\caption{Cross sections of the invariant M$_{p\pi^{+}}$ mass from
the p~p~$\rightarrow$~p~$\pi^{+}$X reaction.
The extracted values are given and are related to the four  
inserts of Fig. 35, justifying the extraction of 
a narrow structure at M=1136~MeV.}
\label{tab5}
\end{table}
\begin{table}
\begin{tabular} [h]{c c c c c c}
Figure 35& Mass& width& S.D.&T$_{p}$&$\theta$\\
 &(MeV)&(MeV)& &(GeV)\\
\hline
 (a)&1143  &7.9 &6.3&2.1  &0.7$^{0}$\\
 (b)&1133.7&10.6&11&2.1&3$^{0}$\\
 (c)&1130.4&11.3&4.9&1.805&3.7$^{0}$\\
 (d)&1130.8&8.6&3.6&1.805&9$^{0}$\\
\end{tabular}
\end{table}
\begin{table}
\caption{Cross sections of the invariant M$_{p\pi^{+}}$ mass from 
the p~p $\rightarrow$ p ${\pi^+}X$ reaction. 
The extracted values are given and are related to the four
inserts of Fig. 36 justifying the extraction of 
a narrow structure at M=1173~MeV.}
\label{tab6}
\end{table}
\begin{table}
\begin{tabular} [h]{c c c c c c}
Figure 36& Mass& width& S.D.&T$_{p}$&$\theta$\\
 &(MeV)&(MeV)& &(GeV)\\
\hline
 (a)&1173.4&8.1&7&1.52&13$^{0}$\\
 (b)&1166.9&4.2&3.4&1.52&0$^{0}$\\
 (c)&1170.4&5.9&2.8&1.52&2$^{0}$\\
 (d)&1176.6&7.8&3.1&1.805&6.7$^{0}$\\
\end{tabular}
\end{table}
\newpage
\begin{table}
\caption{Cross sections of the invariant M$_{p\pi^{+}}$ mass from
the p~p $\rightarrow$ p ${\pi^+}X$ reaction.
The extracted values are given and are related to the four
inserts of Fig. 37 justifying the extraction of
a narrow structure at M=1249~MeV.}
\label{tab7}
\end{table}
\begin{table}
\begin{tabular} [h]{c c c c c c}
Figure 37& Mass& width& S.D.&T$_{p}$&$\theta$\\
 &(MeV)&(MeV)& &(GeV)\\
\hline
(a)&1247.3&13.9&10.4&1.52&0$^{0}$\\
(b)&1251.1&9.5&8.3&1.52&2$^{0}$\\
(c)&1250.7&10.8&7.3&1.52&5$^{0}$\\
(d)&1250.2&5.0&3.0&1.52&9$^{0}$\\
\end{tabular}
\end{table}
\vspace*{1cm}
\begin{table}
\caption{Cross sections of the invariant M$_{p\pi^{+}}$ mass
and of the missing mass $M_{X}$ from
the p~p $\rightarrow$ p ${\pi^+}X$ reaction. 
The extracted values are given and are related to the three
inserts of Fig. 38 justifying the extraction of
a narrow structure at M=1277~MeV. Insert (d) illustrates the observation of
a structure close to M=1330~MeV.}
\label{tab8}
\end{table}
\begin{table}
\begin{tabular} [h]{c c c c c c c}
Figure 38& Mass& width& S.D.&T$_{p}$&$\theta$&observable\\
 &(MeV)&(MeV)& &(GeV)\\
 \hline
 (a)&1282.8&8.3&3.5&1.805&6.7$^{0}$&$M_{p\pi^{+}}$\\
 (b)&1270.0&8.1&3.2&1.805&3.7$^{0}$&$M_{X}$\\
 (c)&1277.0&6.6&2.4&2.1&9$^{0}$&$M_{p\pi^{+}}$\\
 (d)&1327.1&5.3&2.7&2.1&9$^{0}$&$M_{X}$\\
\end{tabular}
\end{table}
\newpage
\begin{table}
\caption{Cross sections of the missing mass $M_{X}$ from
the p~p~$\rightarrow$~p~${\pi^+}$X and d~p~$\rightarrow$~p~p~X 
reactions. 
The extracted values are given and are related to the two
inserts of Fig. 39 justifying the possibility of the extraction of
a narrow structure close to M=1384~MeV.}
\label{tab9}
\end{table}
\begin{table}
\begin{tabular} [h]{c c c c c c c}
Figure 39& Mass&reaction&width& S.D.&T$_{p}$&$\theta$\\
&(MeV)& &(MeV)& &(GeV)\\
\hline
(a)&1375.3&p~p~$\rightarrow$~p~${\pi^+}$X &12.1 &3.7&1.805&6.7$^{0}$\\
(b)&1391.9&d~p~$\rightarrow$~p~p~X&2.8 &2.6&2.1&17$^{0}$\\
\end{tabular}
\end{table}
\begin{table}
\caption{Experiments whose results are displayed in Fig. 40. The last
column (j) of Fig. 40 shows also one mass (M=1136~MeV) extracted from the
np$\to$pX experiment from LAMPF [13].}
\label{tab10}
\end{table}
\begin{table}
\begin{tabular} [h]{c c c c c}
column &reaction&variable&incident energy (MeV)&reference\\
\hline
(a)&pp$\to \pi^{+}$pX&$M_{X}$&1520&this work\\
(b)&pp$\to \pi^{+}$pX&$M_{X}$&1805&this work\\
(c)&pp$\to \pi^{+}$pX&$M_{X}$&2100&this work\\
(d)&pp$\to \pi^{+}$pX&$M_{p\pi}$&1520&this work\\
(e)&pp$\to \pi^{+}$pX&$M_{p\pi}$&1805&this work\\
(f)&pp$\to \pi^{+}$pX&$M_{p\pi}$&2100&this work\\
(g)&dp$\to$ppX&$M_{X}$&1722&this work\\
(h)&dp$\to$ppX&$M_{X}$&2100&this work\\
(i)&$\gamma n\to p\pi^{+}\pi^{0}$&$M_{p\pi}$&500-800&[18] (MAMI)\\
(j)&$\gamma p\to\pi^{+}$n&$W_{CM}$&300-2000&[19 ] (Bonn)\\
\end{tabular}
\end{table}


\begin{thebibliography}{99}
\bibitem{str} The structure at M=1.044~GeV is rather weakly excited
at T$_{p}$=1.52~GeV, $\theta$=2$^{0}$.
\end{thebibliography}
\end{document}